\documentclass{epl}

\usepackage{epsfig}

\title{Oscillatory instability in a driven granular gas}
\author{Evgeniy Khain \and Baruch Meerson}
\institute{Racah Institute of Physics, Hebrew University of
Jerusalem, Jerusalem 91904, Israel}


\begin{document}

\maketitle

\begin{abstract}
We discovered an oscillatory instability in a system of
inelastically colliding hard spheres, driven by two opposite
``thermal" walls at zero gravity.  The instability, predicted by a
linear stability analysis of the equations of granular
hydrodynamics, occurs when the inelasticity of particle collisions
exceeds a critical value. Molecular dynamic simulations support
the theory and show a stripe-shaped cluster moving back and forth
in the middle of the box away from the driving walls. The
oscillations are irregular but have a single dominating frequency
that is close to the frequency at the instability onset, predicted
from hydrodynamics.
\end{abstract}

\section{Introduction}

Granular gas - a system of inelastically colliding hard spheres -
is a minimalistic model of granular flow that has received much
recent attention \cite{review}. In particular, this model captures
\textit{clustering}, a generic and fascinating phenomenon in rapid
granular flows resulting from the collisional loss of the kinetic
energy of random motion of particles. The formation and structure
of granular clusters have been investigated in many works, both in
the context of a freely "cooling" granular gas
\cite{freelycooling} and for driven granular gases
\cite{Kadanoff2,Kudrolli,Grossman,Esipov,Urbach,Brey,Demon,Tobochnik,LMS,Brey2,KM,Argentina,LMS2,MPSS,KMS,Kudrolli1}.
The basic physics of clustering can often be described in a
hydrodynamic language, in terms of a collective condensation mode
driven by bulk losses of energy. Similar condensation processes,
driven by \textit{radiative} energy losses in gases and plasmas,
have been known for a long time \cite{Field}. A complete
understanding of the properties of granular gas (a system
intrinsically far from equilibrium) is still lacking. One
important open question is the exact criteria for the validity of
kinetic theory and Navier-Stokes hydrodynamics, developed in the
80-ies \cite{hydro,Jenkins}.

\textit{Driven} granular gas has the advantage that a steady state
is achievable: the energy supplied into the system can be balanced
by the collisional energy losses. We have focused in this work on
a simple model: inelastic smooth hard spheres confined in a
rectangular box and driven by two opposite ``thermal" walls at
zero gravity. Prototypical granular systems of this type were
considered by many workers. In the early work, theoretical
~\cite{Grossman,Brey,Tobochnik} and experimental~\cite{Kudrolli},
the ``stripe state": a denser and ``colder" stripe-like cluster of
particles, away from the driving wall(s), was documented. The
formation of the stripe state has a simple explanation. Due to the
inelastic particle collisions, the granular temperature goes down
with an increase of the distance from the thermal wall(s). To
maintain the steady-state pressure balance, the particle density
should increase with this distance. For a large density contrast,
the stripe state is observed.

It has been found more recently that the stripe state is only one,
trivial state of this system.  In a certain region of parameters
the stripe state becomes either unstable or metastable and
undergoes a spontaneous symmetry breaking and phase separation. As
the result, a higher-density "phase" coexists with a lower-density
phase in the direction \textit{parallel} to the driving
wall(s)~\cite{LMS,Brey2,KM,LMS2,MPSS,KMS,Argentina}. A fascinating
phenomenology of this far-from-equilibrium phase separation, and
analogy with van der Waals gas~\cite{Argentina} put this system
into a list of pattern-forming systems far from equilibrium.

In this work we report a new and very different symmetry-breaking
instability in the same system. Like the phase separation
instability, the new instability occurs when the inelasticity
$q=(1-r)/2$ exceeds a critical value depending on the rest of the
parameters of the system (here $r$ is the coefficient of normal
restitution of the particle collisions). In contrast to the phase
separation instability, this is an instability in the direction
\textit{normal} to the driving wall(s), and it is oscillatory. As
it develops, the stripe-shaped cluster in the middle of the system
exhibits large irregular oscillations with a single dominating
frequency.

The rest of the paper is organized as follows. First, we formulate
a hydrodynamic model and briefly describe the static stripe state.
Then we present a linear hydrodynamic stability analysis of the
stripe state that predicts oscillatory instability. The
instability is then observed in molecular dynamic (MD)
simulations. Finally, we briefly discuss our results.

\section{Model and static stripe state}

Let $N \gg 1$ identical smooth hard disks with diameter $d$ and
mass $m$ move and inelastically collide inside a two-dimensional
rectangular box with width $L$ and height $H$. There is no gravity
in the model. The system is driven by two thermal walls, with the
same temperature $T_0$, located at $x=-L/2$ and $x=L/2$. We assume
nearly elastic particle collisions, $q \ll 1$, and moderate
granular densities: $n/n_c \leq 0.5$, where $n$ is the local
number density of the particles, and $n_c = 2/(\sqrt{3} d^2)$ is
the hexagonal dense packing density. These assumptions enable us
to employ a standard version of Navier-Stokes granular
hydrodynamics~\cite{Jenkins}. Hydrodynamics is expected to be
valid when the mean free path of the particles is much smaller
than any length scale (and the mean time between two consecutive
collisions is much smaller than any time scale) described
hydrodynamically. Let us measure the distance in the units of $L$,
the time in the units of $L/T_0^{1/2}$, the density
$n(\mathbf{r},t)$ in the units of $n_c$, the granular temperature
$T(\mathbf{r},t)$ in the units of $T_0$, and the mean velocity
$\mathbf{v}(\mathbf{r},t)$ in the units of $T_0^{1/2}$. Then the
hydrodynamic equations~\cite{Jenkins,corrections} become
dimensionless:
$$
d n / d t + n\, {\mathbf \nabla} \cdot {\mathbf v} = 0\,, \;\;\;
\;\; n \, d{\mathbf v}/dt = {\mathbf \nabla} \cdot {\mathbf P} \,,
$$
\begin{equation}
n\, dT/dt + p\, {\mathbf \nabla} \cdot {\mathbf v} = \epsilon\,
{\mathbf \nabla} \cdot (T^{1/2} F_1 {\mathbf \nabla} T) -
\,\epsilon\, R\, n\, G\,
   T^{3/2}\,.
\label{hydroeqns}
\end{equation}
Here ${\mathbf P} =\left[ - p + \epsilon n G T^{1/2}{\rm tr}\,
({\mathbf D}\,) \right] {\mathbf I} + \epsilon\, F_2 T^{1/2}\,
\hat{{\mathbf D}} \,$ is the stress tensor, $p = n T (1+2G)$ is
the equation of state, ${\mathbf D} = (1/2)\left[{\mathbf \nabla
v} + ({\mathbf \nabla v})^{T}\right]$ is the rate of deformation
tensor, $\hat{{\mathbf D}}={\mathbf D}-(1/2)\, {\rm tr}\,
({\mathbf D}\,)\, {\mathbf I}$ is the deviatoric part of ${\mathbf
D}$, and ${\mathbf I}$ is the identity tensor. The functions $F_1$
and $F_2$ are the following:
\begin{eqnarray}
F_1=nG\left[1+\frac{9\pi}{16}\left(1+\frac{2}{3G}\right)^2\right]
\;\;\; \mbox{and} \;\;\;
F_2=nG\left[1+\frac{\pi}{8}\left(1+\frac{1}{G}\right)^2\right]\,,\nonumber
\label{E}
\end{eqnarray}
where $$ G=\frac{\pi\,n
}{2\sqrt{3}}\frac{\left(1-\frac{7\pi\,n}{32\sqrt{3}}\right)}{\left(1-\frac{\pi\,n}{2\sqrt{3}}\right)^2}.
$$ The two scaled parameters entering Eqs.
(\ref{hydroeqns}) are the small parameter $\epsilon=2\pi^{-1/2} d
/ L \ll 1$ and the parameter $R=(16 / \pi) (1-r) \epsilon^{-2}$
that shows the relative role of the collisional heat loss and heat
conduction. The boundary conditions for the temperature become
$T(x=-1/2, y, t)=T(x=1/2, y, t)=1$. For the velocity we demand
zero normal components and slip (no stress) conditions at all
boundaries. We assume that $H$ is small enough, so that no
symmetry breaking can occur in the $y$-direction
\cite{LMS,Brey2,KM}. Therefore, the hydrodynamic variables are
independent of $y$, while the mean flow, if any, is directed along
the $x$-axis. The conservation of the total number of particles
yields $ \int_{-1/2}^{1/2} dx\, n(x,t) = f\,, \label{conservation}
$ where $f=N / (L H n_c)$ is the average area fraction of the
grains, an additional scaled parameter of the problem.

\begin{figure}[ht]
\begin{tabular}{cc}
\epsfxsize=7.0 cm \epsffile{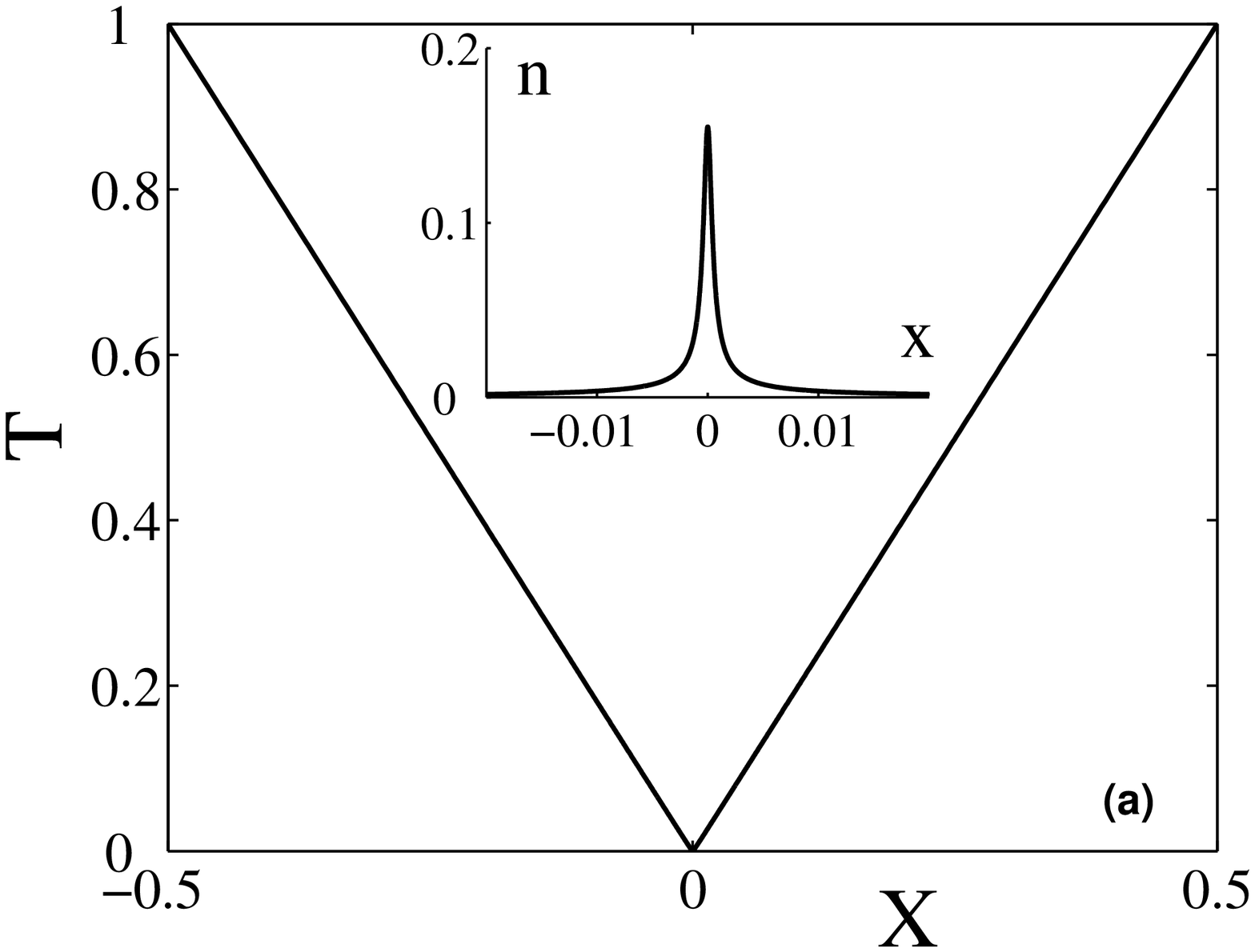} &\hspace{-0.2cm}
\vspace{-0.3 cm} \epsfxsize=7.0 cm \epsffile{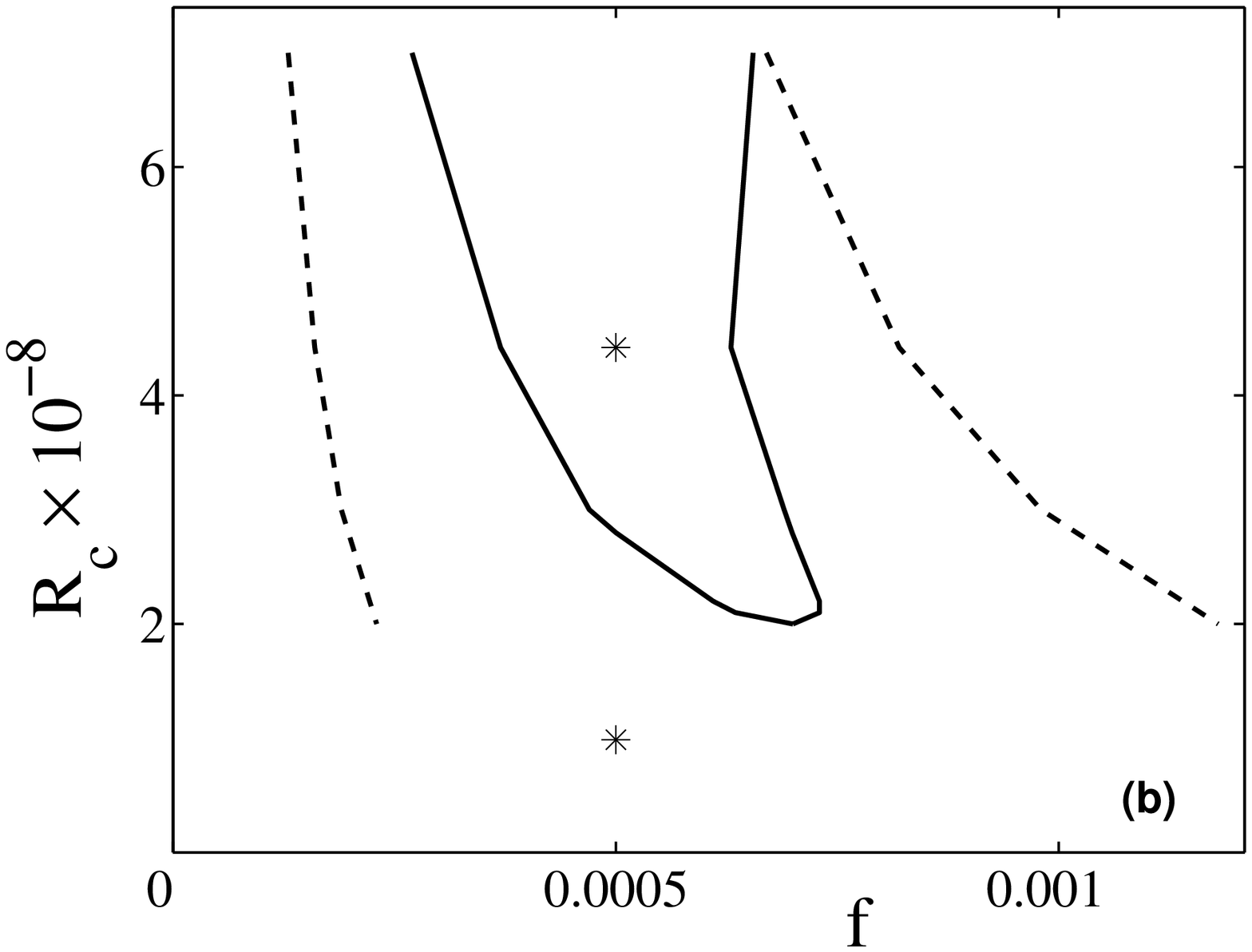}
\end{tabular}
\caption{(a) - The scaled temperature profile of the static stripe
state for $R=4.42\cdot 10^8$ and $f=0.00063$. The inset shows the
respective density profile. Except in a small vicinity of $x = 0$,
the gas is dilute in this example. As the result, the scaled
temperature profile there is $T \simeq 2 |x| $, as follows from
the dilute limit of Eqs. (\ref{steadystate}). (b) - The solid line
is the oscillatory instability threshold $R=R_c$ versus the
average area fraction $f$ at $\epsilon=4\cdot 10^{-5}$. The
instability occurs when $R>R_c$. The dashed lines show the borders
of the spinodal interval of the phase separation instability that
develops in the same system at large enough $H$. The two asterisks
correspond to two MD simulations (see Figs. \ref{snapshot1} -
\ref{power}).} \label{hydr}
\end{figure}

The static stripe state is described by the equations
\begin{equation}
\left[n_s T_s (1+2G_s)\right]^{\prime}=0 \;\;\; \mbox{and} \;\;\;
(T_s^{1/2} J_1 T_s^{\prime})^{\prime}=R\, n_s G_s\,T_s^{3/2},
\label{steadystate}
\end{equation}
where $G_s=G(n_s)$, $J_1=F_1(n_s)$, and the primes denote the
$x$-derivatives. Figure \ref{hydr}a shows an example of the
temperature and density profiles of this state. Notice that,
despite the presence of relatively large temperature and density
gradients in the stripe state, the mean free path of the particles
remains, at $q \ll 1$,  much smaller than the characteristic
length scale of these gradients \cite{Grossman}. Therefore,
hydrodynamics remains valid.

\section{Linear stability analysis}
The linear stability analysis involves linearization of Eqs.
(\ref{hydroeqns}) around the static profiles $n_s(x)$ and
$T_s(x)$. We substitute
\begin{equation}
n(x,t)= n_s(x)+{\rm e}^{-i \omega t}\, \nu(x), \;\;\; T(x,t)=
T_s(x)+{\rm e}^{-i \omega t}\, \Theta (x), \;\;\; \mbox{and}
\;\;\; v(x,t)= {\rm e}^{-i \omega t}\, \psi(x) , \label{modes}
\end{equation}
where $\mbox{Im} \,\omega>0$ ($<0$) corresponds to growth
(damping) of the small perturbation, and obtain the linearized
equations:
\begin{eqnarray}
\label{lin1} &-&i\omega \nu + (n_s \psi)^{\prime}=0\,,  \nonumber
\\
&-&i\omega n_s \psi = - \left(n_s (1+2 \, G_s) \, \Theta
\right)^{\prime} - (T_s \, J_3 \, \nu )^{\prime} +
\epsilon\left(T_s^{1/2} \, n_s \, G_s \, \psi^{\prime}
\right)^{\prime} + \frac{\epsilon}{2}\left( T_s^{1/2} \, J_2 \,
\psi^{\prime} \right)^{\prime}\,,  \nonumber
\\
&-& i\omega \, n_s \, \Theta + n_s \, \psi \, T_s^{\prime} + n_s
\, T_s \, (1+2 \, G_s) \, \psi^{\prime}   \nonumber
\\ &=& \epsilon \left[ (T_s^{1/2} \, \Theta)^{\prime} \, J_1 \right]^{\prime}
+ \epsilon \left( T_s^{1/2} \, T_s^{\prime} \, J_4 \,
\nu\right)^{\prime} - \epsilon \, R \, \left( \frac{3}{2} \,
T_s^{1/2} \, n_s \, G_s \, \Theta + T_s^{3/2} \, J_5 \, \nu
\right),
\end{eqnarray}
where $J_2=F_2(n_s)$, $ J_3=1+2G_s+2n_s d G_s/d n_s $,
\begin{eqnarray}
J_4 = \frac{1}{16 G_s} \left[ 4 \pi + 12 \pi G_s + (16 + 9 \pi )
G_s^2 \right] + \frac{n_s}{16 G_s^2} \, \left[ - 4 \pi + ( 16 + 9
\pi ) G_s^2 \right] \, \frac{d G_s}{d n_s}\,, \nonumber
\end{eqnarray}
$J_5=G_s+n_s d G_s/d n_s $ and $d G_s/d
n_s=(9\pi/4)(16\sqrt{3}+\pi n_s)(6-\sqrt{3} \pi n_s)^{-3}$.
Eliminating $\nu(x)$ from equations (\ref{lin1}), we arrive at a
linear eigenvalue problem for the two-component eigenvector
${\mathbf U} (x) = \left[ \psi(x),\,\Theta(x) \right]$,
corresponding to the complex eigenvalue $\omega$:
\begin{equation}
{\mathbf A}\, {\mathbf U}^{\prime \prime} + {\mathbf B}\, {\mathbf
U}^{\prime} + {\mathbf C}\, {\mathbf U} = 0\,.
\label{vectorsystem}
\end{equation}
The elements of matrices $\mathbf{A}$, $\mathbf{B}$ and
$\mathbf{C}$ are
\begin{eqnarray}
A_{11}&=&-(n_s/i\omega) \, T_s \, J_3 + \epsilon \, a_0 \, n_s \,
G_s + (\epsilon/2) \, a_0 \, J_2, \;\;\; A_{21}= \epsilon \, (n_s
/ i\omega) \, T_s^{1/2} \, T_s^{\prime} \, J_4, \nonumber
\\
A_{12}&=&0, \;\;\; A_{22}= \epsilon \, a_0 \, J_1, \nonumber
\\
B_{11}&=& -(n_s / i\omega) \, T_s^{\prime} \, J_3 - (n_s /
i\omega) \, T_s \, n_s^{\prime} \, J_6 - (2 n_s^{\prime} /
i\omega) \, T_s \, J_3 + \epsilon \, a_1 \, n_s \, G_s + \epsilon
\, a_0 \, n_s^{\prime} \, J_5
 \nonumber \\
&+& (\epsilon/2) \, a_1 \, J_2 + (\epsilon/2) \, a_0 \,
n_s^{\prime} \, J_7 , \nonumber
\\
B_{21}&=& -n_s \, T_s \, (1+2 \, G_s) + \epsilon \, (n_s /
i\omega) \, ( T_s^{1/2} \, T_s^{\prime})^{\prime} \, J_4 +
\epsilon \, (n_s / i\omega) \, T_s^{1/2} \, T_s^{\prime} \,
n_s^{\prime} \, J_8 \nonumber \\ &+& \epsilon \, (2 n_s^{\prime} /
i\omega) \, T_s^{1/2} \, T_s^{\prime} \, J_4 - \epsilon \, R \,
(n_s / i\omega) \, T_s^{3/2} \, J_5, \nonumber
\\
B_{12}&=& - n_s \, (1+2 \, G_s), \;\;\; B_{22}= 2 \, \epsilon \,
a_1 \, J_1 + \epsilon \, a_0 \, n_s^{\prime} \, J_4, \nonumber
\\
C_{11} &=& i\omega \, n_s - (n_s^{\prime} / i\omega) \,
T_s^{\prime} \, J_3 - ((n_s^{\prime})^2 / i\omega) \, T_s \, J_6 -
(n_s^{\prime\prime} / i\omega) \, T_s \, J_3 , \nonumber
\\
C_{21}&=& -n_s \, T_s^{\prime} +  \epsilon \, (n_s^{\prime} /
i\omega) \, ( T_s^{1/2} \, T_s^{\prime})^{\prime} \, J_4 +
\epsilon \, (n_s^{\prime} / i\omega) \, T_s^{1/2} \, T_s^{\prime}
\, n_s^{\prime} \, J_8 \nonumber \\ &+& \epsilon \,
(n_s^{\prime\prime} / i\omega) \, T_s^{1/2} \, T_s^{\prime} \, J_4
- \epsilon \, R \, (n_s^{\prime} / i\omega) \, T_s^{3/2} \, J_5,
\nonumber
\\
C_{12}&=& - n_s^{\prime} \, J_3, \;\;\; C_{22} = i\omega \, n_s +
\epsilon \, a_2 \, J_1 + \epsilon \, a_1 \, n_s^{\prime} \, J_4 -
(3/2) \, \epsilon \, R \, n_s \, G_s \, T_s^{1/2},
\end{eqnarray}
and
\begin{eqnarray}
J_6&=&4 \, d G_s/d n_s  +2 \, n_s \, d^2 G_s/d n_s^2, \nonumber
\\
J_7&=& (1/8) \, G_s^{-1} \left[ \pi + 2 \pi G_s + (8 + \pi) G_s^2
\right] + (1/8) \, G_s^{-2} n_s \left[ - \pi + (8 + \pi) G_s^2
\right] d G_s/d n_s , \nonumber
\\
J_8&=& (\pi/2) \, G_s^{-3} \, n_s \, (d G_s/d n_s)^2
 - (\pi/4) \, G_s^{-2} \, (2 \, d G_s/d n_s + n_s \, d^2 G_s/d n_s^2)
\nonumber \\ &+& (1/16) (16 + 9 \pi) \, (2 \, d G_s/d n_s + n_s \,
d^2 G_s/d n_s^2 ), \nonumber \\
\frac{d^2 G_s}{d n_s^2} &=& (9 \pi^2 / 2) \, (75 \, + \, \sqrt{3}
\,\pi \, n_s) \, (6 \, - \, \sqrt{3} \, \pi \, n_s)^{-4}\,.
\nonumber
\end{eqnarray}
We have denoted for brevity $a_0 = T_s^{1/2}(x)\,,\,  a_1 =
a_0^{\prime}$ and $a_2 = a_0^{\prime\prime}$. The boundary
conditions for the functions $\Theta$ and $\psi$ are the
following:
\begin{equation}
\Theta (\pm1/2)=  \psi (\pm1/2)  = 0\,. \label{boundcond}
\end{equation}
Equations (\ref{vectorsystem})-(\ref{boundcond}) define a linear
\textit{boundary-value} problem: there are two boundary conditions
at one wall, and two at the other wall. A simple and accurate
numerical algorithm, that we developed earlier \cite{conv},
enables one to avoid the unpleasant shooting in two parameters.
The algorithm yields the complex value of $\omega$. Varying $R$ at
fixed $f$ and $\epsilon$, we determined the critical value $R=R_c$
for the instability onset from the condition ${\rm Im} \,\omega =
0$. The instability occurs at $R>R_c$. Given the rest of the
parameters, the instability occurs when $q$ exceed a critical
value. Importantly, the instability is \textit{oscillatory}, as
${\rm Re} \,\omega \neq 0$ at the onset. Figure \ref{hydr}b shows
the critical value $R=R_c$ versus the average area fraction $f$ at
a fixed $\epsilon$.

\section{MD simulations}

To verify the hydrodynamic predictions, and to follow the dynamics
of an oscillating stripe in a nonlinear regime,  we performed MD
simulations with $N=3390$ inelastic hard disks with unit mass and
diameter. As Fig. \ref{hydr}b shows, the oscillatory instability
region lies within the spinodal interval of the phase separation
instability \cite{LMS,KM,Argentina}. The phase separation
instability is suppressed when the aspect ratio $H/L$ is less than
some threshold value depending on $R$ \cite{LMS,Brey2,KM}. It has
been found recently that the system exhibits large fluctuations
even well below the threshold \cite{MPSS}. Therefore, to isolate
the oscillatory instability from the phase separation instability,
we worked with a very small aspect ratio: the box dimensions were
$H = 208$ and $L = 28\, 209$. The parameters  $f=5\cdot 10^{-4}$
and $\epsilon=4 \cdot 10^{-5}$ were fixed. The parameter $R$
varied from $4.42 \cdot 10^8$ to $9.55 \cdot 10^6$. This was
achieved by varying $r$ from $0.85$ to $0.997$. We employed the
event-driven algorithm described in the book of Rapaport
\cite{Rapaport}. Upon collision with a thermal wall, the normal
component of the particle velocity was drawn from a Maxwell
distribution with $T_0=1$, while the tangential component of the
velocity remained unchanged. The initial spatial distribution of
the particles was uniform; the initial velocity distribution was
Maxwell's with $T_0=1$.
\begin{figure}
\vspace{0 cm} \center{\epsfxsize=7.0 cm \epsffile{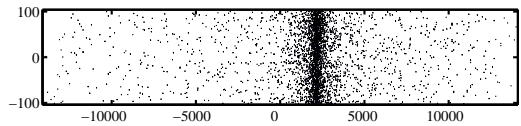}}
\caption{A snapshot of the system of $N=3390$ inelastic hard disks
within the instability region, obtained in a MD simulation. The
left and right walls are thermal walls. The coefficient of normal
restitution $r=0.85$. See the text for the rest of the parameters.
A large deviation of the cluster from the center of the system
$x=0$ is clearly seen.} \label{snapshot1}
\end{figure}
\begin{figure}
\begin{tabular}{cc}
\epsfxsize=6.3 cm \epsffile{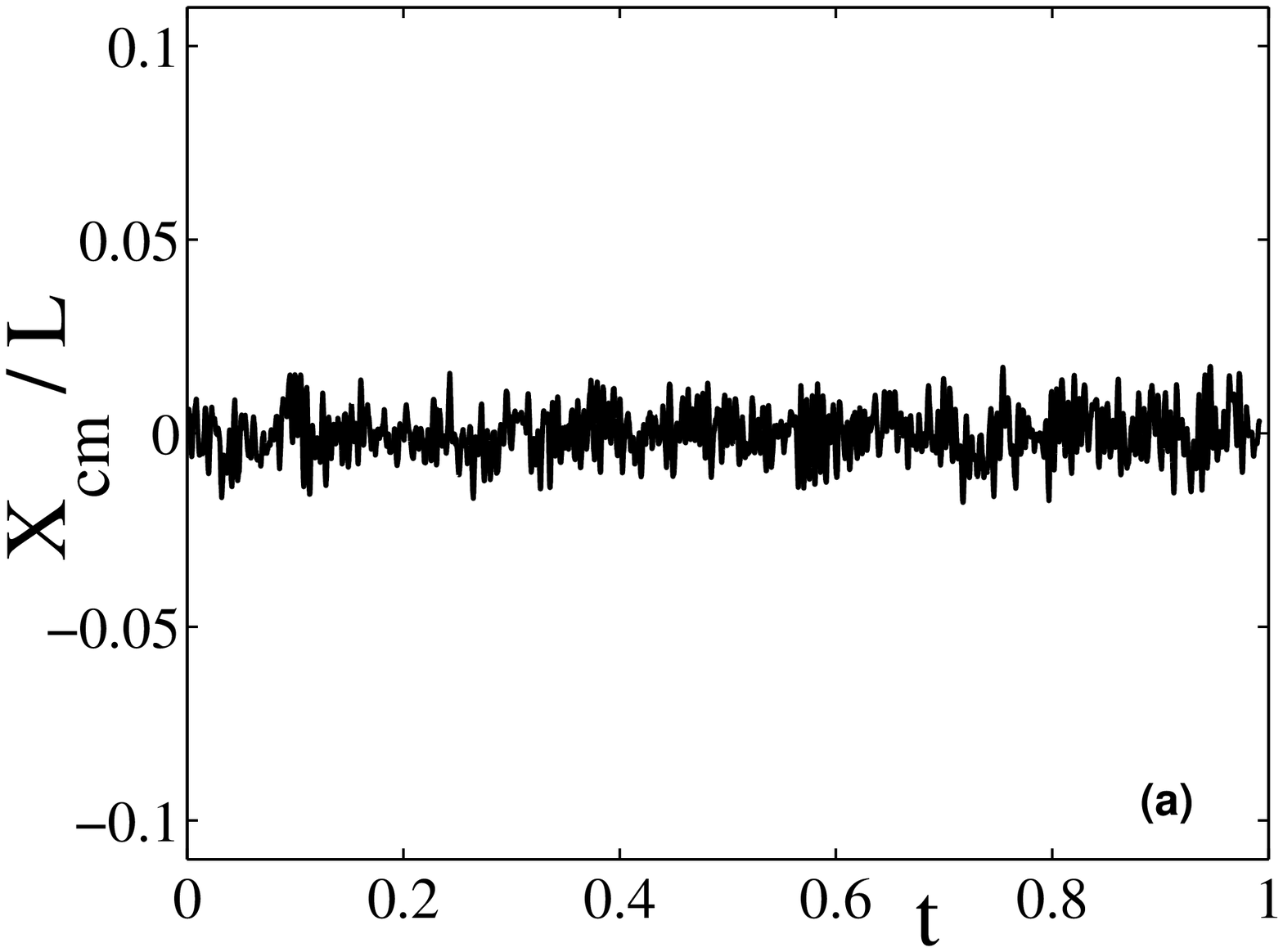} &\hspace{-0.2cm}
\vspace{-0.3 cm} \epsfxsize=6.3 cm \epsffile{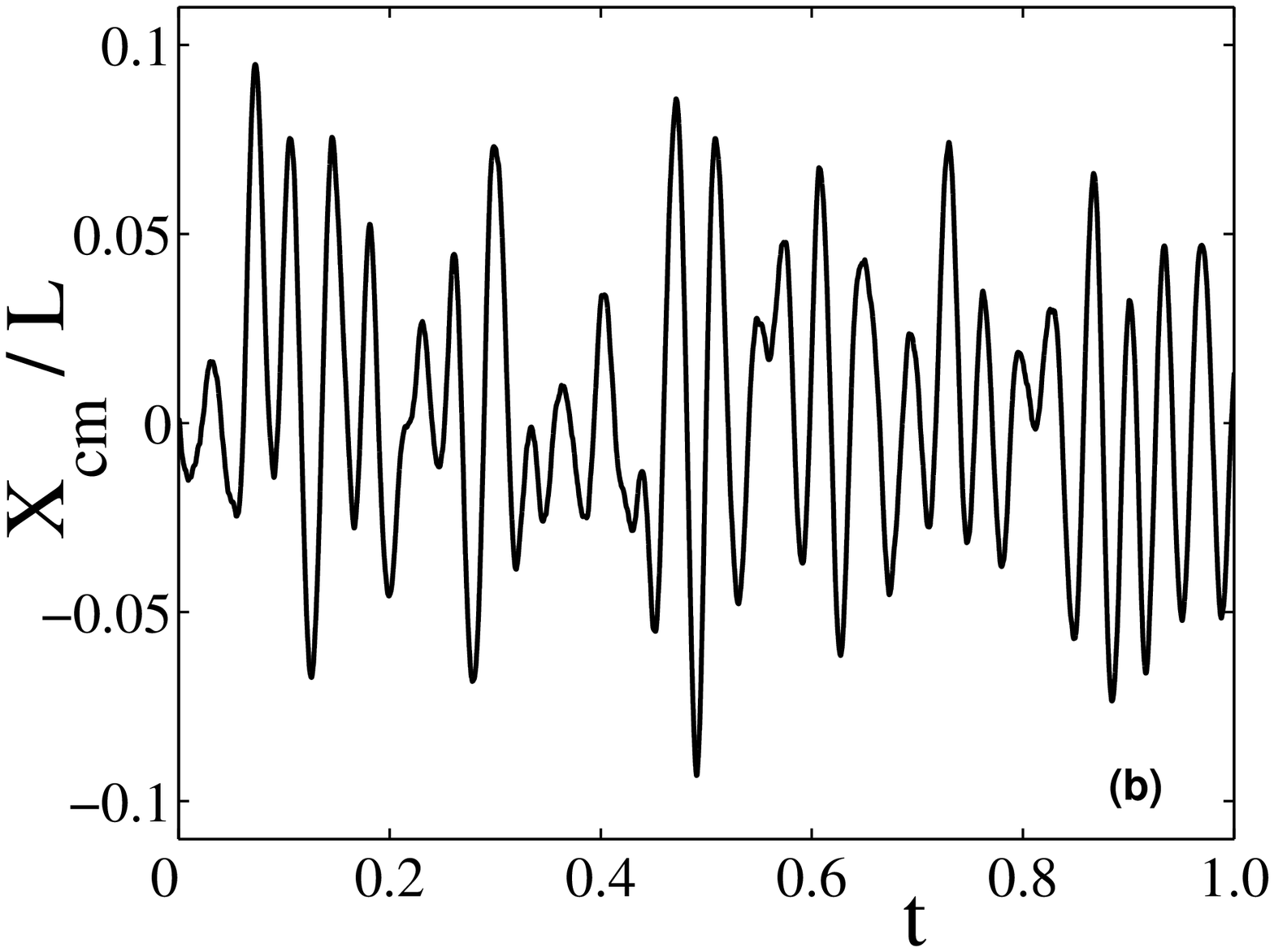}
\end{tabular}
\caption{The time dependence of the center-of-mass coordinate
$X_{cm}$, obtained in MD simulations with $N=3390$ particles
within (a) stable region, $r=0.997$ and (b) unstable region,
$r=0.85$. The time unit is $10^7\,t_{MD}$, where $t_{MD} =
d/T_0^{1/2} = 1$. } \label{xcm}
\end{figure}
Figure \ref{snapshot1} shows a typical snapshot of the system
obtained in a MD simulation within the instability region. A large
deviation of the cluster from the center  of the system $x=0$ is
clearly seen. A movie of this simulation can be seen at
http://huji-phys.phys.huji.ac.il/staff/Khain/abstract.html. The
oscillatory motion of the cluster was monitored by following the
$x$-component of the center of mass of all particles, $X_{cm}$,
versus time. Figure \ref{xcm} shows this dependence as observed in
the stable (a) and unstable (b) regions. The large-amplitude
low-frequency irregular oscillations in the unstable case are
clearly distinguishable from the small-amplitude high-frequency
noise observed in the stable region. The power spectra for these
two cases are shown in Fig. \ref{power}. Notice the large
difference in scale in the horizontal axes and the huge difference
in the vertical axes. In contrast to the low-amplitude noise of
Fig. \ref{power}a, the power spectrum within the instability
region (Fig. \ref{power}b) has a sharp peak at a single frequency
$1.03 \cdot 10^{-4}$. This frequency is fairly close (within $30
\%$) to the hydrodynamic frequency at the instability onset which,
in the units of $t_{MD}^{-1}=1$, is $1.45 \cdot 10^{-4}$. These
two frequencies are not expected to \textit{coincide}. Firstly,
the steady-state oscillations, observed in the MD simulation
(Figs. \ref{xcm}b and \ref{power}b), must be affected by
hydrodynamic nonlinearities that are neglected in the linear
stability analysis. Secondly, the next-order corrections in $q$
can become significant at the moderate value of $r=0.85$ used in
this simulation \cite{corrections}.

The presence of a significant continuum part in the power spectrum
of $X_{cm}(t)$ in Fig. \ref{power}b clearly shows that the cluster
oscillations are chaotic. The same conclusion follows from the
analysis of the phase trajectory in the phase plane $(X_{cm},
\dot{X}_{cm})$. The chaotic component of the oscillations is due
to irregular \textit{cluster} motion, not due to the dynamics of
the surrounding gas. We double-checked it by following the
position $X_{*}$ of the maximum number density of the system
(integrated over the $y$-direction) versus time. The graph of
$X_{*}(t)$ almost coincides with the graph of $X_{cm} (t)$; the
respective power spectra are almost indistinguishable. Obviously,
the chaotic component of the cluster oscillations is completely
missed by the linear stability analysis. Furthermore,
hydrodynamics predicts, at fixed $f$ and $\epsilon$, a sharp
instability onset at $R=R_c$. On the contrary, a fairly smooth
transition is observed in MD simulations. We believe it is due to
the relatively small number of particles that causes significant
fluctuations and smoothes the transition \cite{Haken}. These
fluctuations make it difficult to verify the reentrant behavior
(see the right branch of the solid line in Fig. \ref{hydr}b)
predicted by hydrodynamics.

\begin{figure}
\begin{tabular}{cc}
\epsfxsize=5.9 cm \epsffile{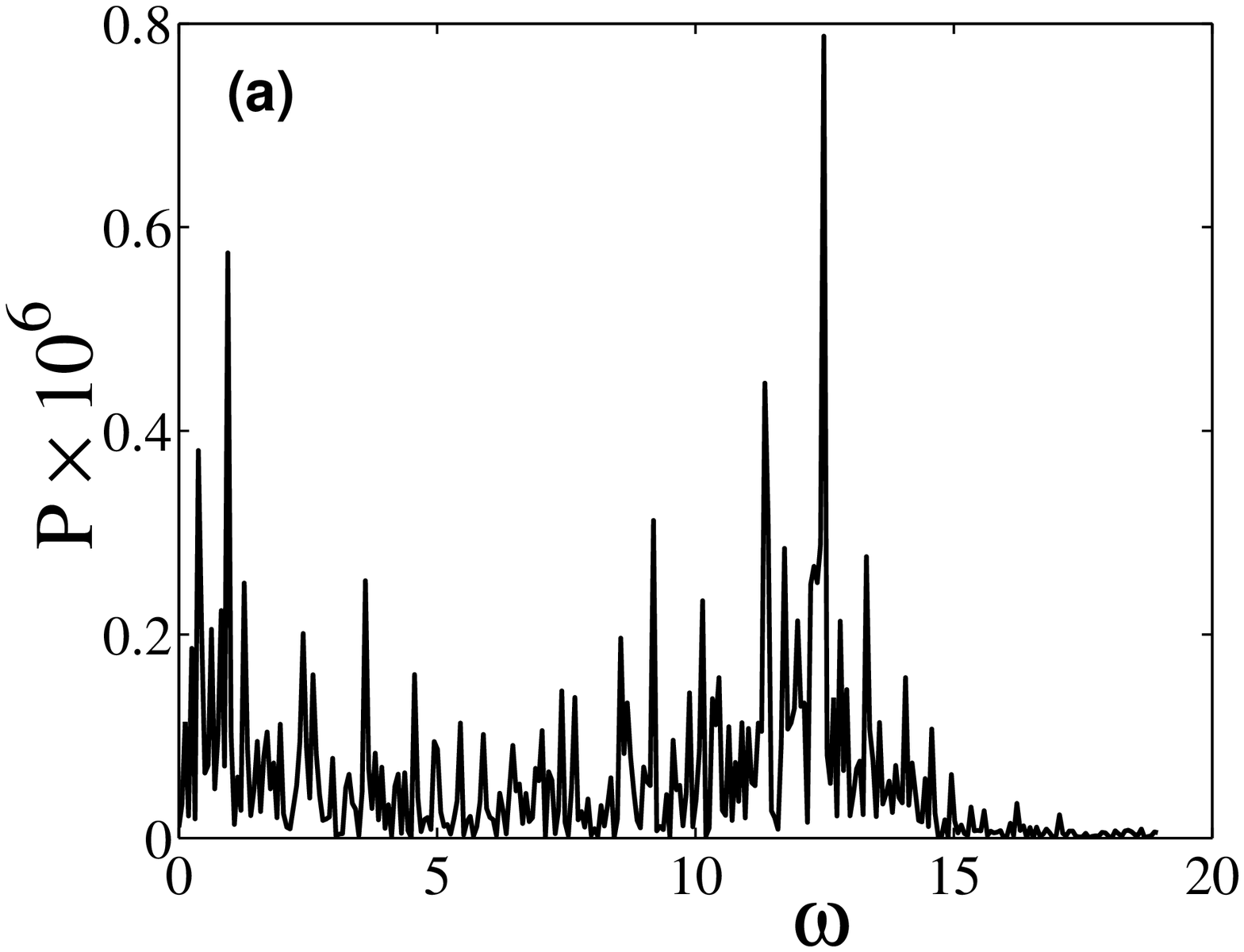} &\hspace{-0.2cm}
\vspace{-0.3 cm} \epsfxsize=5.9 cm \epsffile{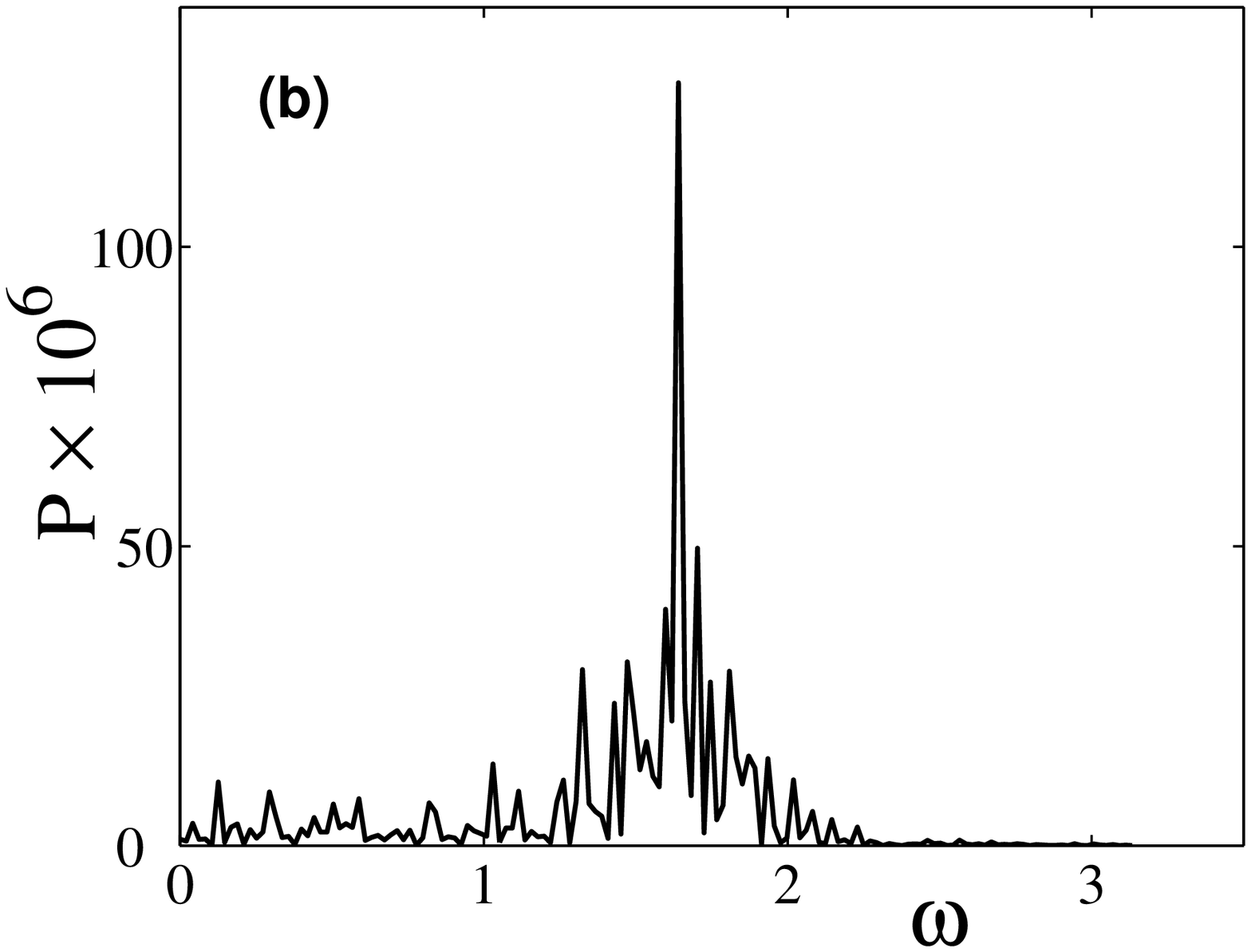}
\end{tabular}
\caption{The power spectra within the stable (a) and unstable (b)
regions. The frequency unit is $2\pi \cdot 10^{-5}$. The
parameters in Figs. a and b are the same as in Figs. \ref{xcm}a
and \ref{xcm}b, respectively. Notice the large differences in
scales.} \label{power}
\end{figure}

\section{Summary and Discussion}

We have discovered a new oscillatory instability in a system of
inelastically colliding hard spheres, driven by two opposite
``thermal" walls at zero gravity. The instability has a
hydrodynamic character and can therefore be interpreted in terms
of the collective modes of the system. There are three such modes,
coupled due to the inhomogeneity of the unperturbed state of the
system. Two of the modes are oscillatory acoustic modes, the third
one is a non-oscillatory entropy mode (the fourth, shear mode, is
suppressed because of the very small aspect ratio of the box). It
is the acoustic modes that become unstable at $R>R_c$. To some
extent, this instability is similar to the acoustic instability
(the so called "overstability") found in externally heated and
radiatively cooling optically thin plasmas \cite{Field}. The
future work should address the nonlinear regime of the cluster
oscillations and clarify the role of discrete-particle noise
versus the spatio-temporal chaos caused by hydrodynamic
nonlinearities.

\acknowledgments

E.K. is very grateful to D.C. Rapaport for guidance in MD
simulations. We thank P. V. Sasorov for a useful discussion. The
work was supported by the Israel Science Foundation (grant No.
180/02).


\begin{thebibliography}{0}
\bibitem{review} \Book{Granular Gases}\Editor{P\"{o}schel T.
\and Luding S.} \Publ{Springer, Berlin} \Year{2001};
 \Book{Granular Gas
Dynamics} \Editor{P\"{o}schel T. \and Brilliantov N.}
\Publ{Springer, Berlin} \Year{2003}; \Name{Goldhirsch I.}
\REVIEW{Annu. Rev. Fluid Mech.}{35}{2003}{267}.
\bibitem{freelycooling} 
\Name{Goldhirsch I. \and Zanetti G.}
\REVIEW{Phys. Rev. Lett.}{70}{1993}{1619}; \Name{McNamara S. \and
Young W. R.} \REVIEW{Phys. Rev. E}{53}{1996}{5089}; \Name{Deltour
P. \and Barrat J. L.} \REVIEW{J. Phys. I. France}{7}{1997}{137};
\Name{van Noije T. P. C., Ernst M. H., Brito R. \and Orza J. A.
G.} \REVIEW{Phys. Rev. Lett.}{79}{1997}{411}; \Name{Brito R. \and
Ernst M. H.} \REVIEW{Europhys. Lett.}{43}{1998}{497}; \Name{Luding
S. \and Herrmann H. J.} \REVIEW{Chaos}{9}{1999}{673}; \Name{Nie X.
B., Ben-Naim E. \and Chen S. Y.} \REVIEW{Phys. Rev.
Lett.}{89}{2002}{204301}.
\bibitem{Kadanoff2} \Name{Du Y., Li H. \and Kadanoff L. P.} \REVIEW{Phys. Rev. Lett.}{74}{1995}{1268}.
\bibitem{Kudrolli} \Name{Kudrolli A., Wolpert  M. \and Gollub J. P.} \REVIEW{Phys. Rev.
Lett.}{78}{1997}{1383}.
\bibitem{Grossman} \Name{Grossman E. L., Zhou T. and Ben-Naim E.} \REVIEW{Phys. Rev. E}{55}{1997}{4200}.
\bibitem{Esipov} \Name{Esipov S. E. \and P\"{o}schel T.} \REVIEW{J. Stat. Phys.}{86}{1997}{1385}.
\bibitem{Urbach} \Name{Olafsen J. S. \and Urbach J. S.} \REVIEW{Phys. Rev.
Lett.}{81}{1998}{4369}; \Name{Nie X. B., Ben-Naim E. \and Chen S.
Y.} \REVIEW{Europhys. Lett.}{51}{2002}{679}; \Name{Cafiero R.,
Luding S. \and Herrmann H. J.} \REVIEW{Phys. Rev.
Lett.}{84}{2000}{6014}.
\bibitem{Brey} \Name{Brey J. J. \and Cubero D.} \REVIEW{Phys. Rev. E}{57}{1998}{2019}.
\bibitem{Demon} \Name{Eggers J.} \REVIEW{Phys. Rev.
Lett.}{83}{1999}{5322}; \Name{van der Weele K., van der Meer D.,
Versluis M. \and Lohse D.} \REVIEW{Europhys.
Lett.}{53}{2001}{328}.
\bibitem{Tobochnik} \Name{Tobochnik J.} \REVIEW{Phys. Rev. E}{60}{1999}{7137}.
\bibitem{LMS} \Name{Livne E., Meerson B. \and Sasorov P. V.} \REVIEW{Phys. Rev. E}{65}{2002}{021302}.
\bibitem{Brey2} \Name{Brey J. J., Ruiz-Montero M. J., Moreno F. \and Garcia-Rojo R.} \REVIEW{Phys. Rev. E}{65}{2002}{061302}.
\bibitem{KM} \Name{Khain E. \and Meerson B.} \REVIEW{Phys. Rev. E}{66}{2002}{021306}.
\bibitem{Argentina} \Name{Argentina M., Clerc M. G. \and Soto R.} \REVIEW{Phys. Rev. Lett.}{89}{2002}{044301}.
\bibitem{LMS2} \Name{Livne E., Meerson B. \and Sasorov P. V.} \REVIEW{Phys. Rev. E}{66}{2002}{050301(R)}.
\bibitem{MPSS} \Name{Meerson B., P\"{o}schel T., Sasorov P.V. \and Schwager T.} e-print cond-mat/0208286.
\bibitem{KMS} \Name{Khain E., Meerson B. \and Sasorov P. V.} in preparation.
\bibitem{Kudrolli1} \Name{Blair D. L. \and Kudrolli A.} \REVIEW{Phys. Rev.
E}{67}{2003}{041301}.
\bibitem{Field} \Name{Field G. B.} \REVIEW{Astrophys.
J.}{142}{1965}{531}; \Name{Meerson B.} \REVIEW{Rev. Mod.
Phys.}{68}{1996}{215}.
\bibitem{hydro} \Name{Haff P. K.} \REVIEW{J. Fluid
Mech.}{134}{1983}{401}.
\bibitem{Jenkins} \Name{Jenkins J. T. \and
Richman M. W.} \REVIEW{Phys. Fluids}{28}{1985}{3485}.
\bibitem{corrections} A more complete theory should include corrections of the order
of $q$ to the constitutive relations, as well as a
\textit{density} gradient term, also of the order of $q$, in the
heat flux \cite{Brey1}. In the nearly-elastic limit $q \ll 1$ that
we consider in our hydrodynamic analysis, these corrections can be
neglected.
\bibitem{Brey1} \Name{Brey J. J., Dufty J. W., Kim C. S. \and Santos A.}
\REVIEW{Phys. Rev. E}{58}{1998}{4638}.
\bibitem{conv} \Name{Khain E. \and Meerson B.} \REVIEW{Phys. Rev. E}{67}{2003}{021306}.
\bibitem{Rapaport} \Name{Rapaport D. C.} \Book{The Art of Molecular Dynamics Simulation}
\Publ{Cambridge University Press, Cambridge} \Year{1995}.
\bibitem{Haken} \Name{Haken H.} \Book{Synergetics}
\Publ{Springer, Berlin} \Year{1978}.
\end{thebibliography}
\end{document}